# Detecting Volcanism on Extrasolar Planets

L. Kaltenegger[1], W. G. Henning[2] & D. D. Sasselov[1]

[1]*Harvard University, 60 Garden Street, 02138 MA, Cambridge USA*
[2]*Harvard University, EPS, 20 Oxford St, 02138 MA, Cambridge USA*
Email: lkaltene@cfa.harvard.edu

**Abstract**
The search for extrasolar rocky planets has already found the first transiting rocky super-Earth, Corot 7b, with a surface temperature that allows for magma oceans. Here we ask if we could distinguish rocky planets with recent major volcanism by remote observation. We develop a model for volcanic eruptions on an Earth-like exoplanet based on the present day Earth, derive the observable features in emergent and transmission spectra for multiple scenarios of gas distribution and cloudcover. We calculate the observation time needed to detect explosive volcanism on exoplanets in primary as well as secondary eclipse and discuss the likelihood of observing volcanism on transiting Earth to super-Earth sized exoplanets. We find that sulfur dioxide from large explosive eruptions does present a spectral signal that is remotely detectable especially for secondary eclipse measurements around the closest stars and ground based telescopes, and report the frequency and magnitude of the expected signatures. Transit probability of planet in the habitable zone decreases with distance to the host star, making small, close by host stars the best targets.

**Subject headings:** astrobiology, Solar System: Earth, planets and satellites: general, methods: data analysis, techniques: spectroscopic

**1. Introduction**

Recently a large sample of extrasolar planets with masses less than Uranus have been found (see e.g. Mayor et al. 2009, Bouchy 2009) including the first transiting rocky planet CoRoT-7b (Leger et al. 2009) that orbits close enough to its star to possibly have a magma ocean maintained by insolation. The study of planets orbiting nearby stars has entered an era of characterizing massive terrestrial planets (super-Earths). With only two transiting planets in the super-Earth and Mini-Neptune mass range (1-10 Earth masses ($M_E$)) so far detected, these two bodies already indicate the amazing diversity of planets in this mass range that do not exist in our own Solar System. At the same time, planetary volcanism is widespread, varied, and is expected to be a ubiquitous phenomenon on rocky planets. Atmospheric evidence of major eruptions may be one of our earliest pathways to learning about surface processes on such extrasolar rocky worlds. Transmission and emergent (reflection and emission) spectra of terrestrial exoplanets, especially super-Earths that orbit nearby stars, may be observable in the near future with the same technique that has recently proven successful in providing spectra of gas giant exoplanets (see e.g. Grillmair et al. 2008 Swain et al. 2008). Such spectroscopy provides molecular band strengths of multiple detected transitions (in absorption or emission) of a few abundant molecules in the planetary atmosphere, allowing a crude model atmosphere to be constructed, tested and improved. In this paper we develop models for Earth-analogues with present day terrestrial volcanism levels or increased volcanism, and their detectable spectral features, to inform observations with future ground- and space-based facilities like the Extremely Large Telescope (ELT) and the James Webb Space Telescope (JWST). Section 2 describes volcanic outgassing and cases on Earth, section 3 describes our model, section 4 shows our results and section 5 discusses the results.

## 2. VOLCANISM AND CASE STUDIES

### 2.1 Volcanism on Earth

The majority of volcanic gas release on Earth occurs non-explosively near the surface, either as mid-ocean ridge volcanism, fumaroles, or from the venting of low viscosity magmas such as the continuous non-explosive eruption of Kilauea in Hawaii. For exoplanet observations, we are primarily concerned with explosive volcanoes that have the potential to inject material into a planet's stratosphere. Residence times are longer for volcanic gases in the stratosphere due to washout effects at lower altitudes. On Earth, volcanic gases released at or near sea level typically do not reach the stratosphere, as discussed further in section 4.1.

The scenario of $SO_2$ distributed non-explosively into an atmosphere is effectively treated in Kaltenegger and Sasselov, 2009. Therefore this paper focuses primarily on stratospheric gases, and hence large explosive events, which may be investigated by both spectroscopic methods, and even in the presence of thick tropospheric cloudcover.

Explosive, or Plinean volcanism occurs along subduction zones and in continental interiors, where magmas are more silicic and volatile rich. High silica content in continental and older back-arc basin volcanoes leads to high viscosities, the trapping of volatiles as gas bubbles, and a high potential for pressure to build up. These volcanoes, such as Mt. St. Helens, Krakatau, and Mt. Pinatubo, explode with sufficient energy to inject plume gases directly into the stratosphere at altitudes between 12 to 25 km. Less frequent historical explosive super-volcanoes such as Yellowstone, Toba, Mt. Mazama and Tambora have the potential to inject greater gas volumes to even higher altitudes.

Volcanic gas emissions for different types of volcanoes are listed in Table 1 in mole percent by total erupted volume. Emissions principally consist of $H_2O$, $H_2$, $CO_2$, $SO_2$, and $H_2S$; the noble gases He, Ar, and Ne; halogens such as HCl, HF and HBr; minor amounts of CO, $CH_4$, OCS, and $CS_2$; and trace amounts of higher sulfur, mercury and organic compounds. These gases are measured by in-situ sampling of vents and inclusions, aircraft measurements, and satellite remote sensing.

### 2.2 $SO_2$ as a Proxy for Volcanism

$SO_2$ is not present in high quantities in an Earthlike stratosphere without a volcanic eruption, and has a variety of discernable absorption lines. Stratospheric sources of $SO_2$ include only explosive eruptions and small amounts by the diffusion of OCS and $CS_2$ from the troposphere (Self et al. 1996). Sea level injections of $SO_2$ are rapidly converted to $H_2SO_4$ aerosols in the presence of water, and are rained out before significant diffusion to the stratosphere can occur (Halmer et al. 2002). $SO_2$ is typically the third or fourth most common volcanic gas, and remains so across a very broad and robust range of volcanic subtypes. Hydrogen sulfide ($H_2S$) has similar properties, but is often erupted in smaller quantities. Halogen compounds such as HCl, HBr and HF are only erupted in limited quantities, compounded by the fact that they react with $H_2O$ during plume accent, and reach the stratosphere in very limited quantities, with 99.99% washing out before they reach the stratosphere. Noble gases are not detectable remotely.

The primary stratospheric sink of both $SO_2$ and $H_2S$ is conversion to $H_2SO_4$ aerosols, which occurs on a timescale of weeks to months. For most eruptions, an order of magnitude less of $H_2S$ reaches the stratosphere from volcanism than $SO_2$, leaving $SO_2$ as our single primary signal for recent exoplanet volcanic activity. In rarer instances $H_2S$ emissions can approach or slightly exceed $SO_2$, as was the case for Mt. St. Helens and Mt. Usu (Giggenbach et al. 1986).. $H_2SO_4$ aerosols and volcanic ash are not discernible with spectroscopy because they block light, similar to clouds.

About 15 to 21 megatons (Mt, $10^9$ kg) of $SO_2$ are produced annually from all non-explosive volcanic sources (Halmer et al. 2002). This is similar to a single large explosive event, but will generate a weaker signal because it is spread out over a full year, and individual release events are subject to rapid (hours to days) washout by reactions with tropospheric $H_2O$. Thus while large explosive events are less frequent, this is partly offset by the longer residence time of volcanic gases in the stratosphere. By concentrating gases in time, they also allow concentrations to exceed our detection thresholds, something that may never occur for non-explosive volcanism, except in extreme cases (see section 4.2).

## 2.3 Case Scenarios

The eruption of Mount Pinatubo in the Philippines in June 1991 is the most recent large explosive eruption with the most complete stratospheric satellite remote sensing record. This event serves as a baseline for our models. Mount Pinatubo injected 17(+/-2) Mt. of $SO_2$ into the stratosphere over the 4 day primary eruption period (Gerlach et al. 1996). Plume altitudes were predominantly between 12-25 km, with a single event reaching 40 km (Self et al. 1999) This led to the subsequent production of about 20 to 30 Mt $H_2SO_4$ aerosol after 28 days, consuming 10 to 15 Mt in $SO_2$ (assuming aerosol droplets are 75 wt% $H_2SO4$ and 25% $H_2O$) (McCormick and Veiga 1992, Hamill et al. 1977). After 170 days all $SO_2$ appears to have been converted to aerosols. Erupted $SO_2$ that was not lifted into the stratosphere was rapidly scavenged by reactions with water. 3 Mt of HCl were erupted, but due to high temperature reactions in the rising plume, 99.99% was washed out (Halmer et al. 2002) and negligible amounts reached the stratosphere. The Pinatubo plume circled the globe in ~15 days, and was distributed globally after 1 year. Ash from the plume settled out of the atmosphere within weeks, leaving $H_2SO_4$ aerosols as the primary cause of long-term changes in atmospheric optical depth (Dutton et al. 1992, 1994). The $H_2SO_4$ aerosols themselves reside in the stratosphere for approximately 1-3 years. Similar eruptions discussed in section 2 (and in Halmer et al. 2002) suggest that 17 +/-2 Mt $SO_2$ is representative.

The vertical distribution of volcanic gases is mainly controlled by the style of eruption. We define the following two cases for the distribution of bulk $SO_2$ injected from a volcanic eruption into a stratosphere. A single main explosion may place gases into a tighter vertical region (Case 1). A complex eruption consisting of multiple events over multiple days or hours will tend to distribute gases into a broader vertical range, as happened in the ~13km thick case of Mount Pinatubo (Case 2). Case 1 assumes a global distribution of volcanic gasses in a shell of 2 km thickness starting at an altitude of 12km. Case 2 assumes a global distribution in a shell between 12 and 25km. Non-global distributions may occur under ~15 days, however this may not have a strong impact on emergent signal strength because we observe disk integrated spectra. For transmission spectroscopy there may potentially be stronger signals, but only in rare cases where limb orientation effects are favorable and the signal to noise ratio, (SNR) is sufficiently high (see section 4.3) to overcome shortened integration times.

## 2.4 Frequency and Scale of Eruptions

To assess the probability of potentially observing an explosive eruption on an extrasolar Earth-analogue, we estimate eruption frequencies from data on Earth. Table 2 shows the frequency of volcanic events through Earth's history versus their magnitude. Eruption strengths are classified using either the Volcanic Explosivity Index (VEI), which is based mainly on a log scale of erupted volume (Newhall and Self 1982), or using a magnitude scale based on erupted mass (Pyle, 2000) $M = \log(m)-7$, where $m$ is the ejecta mass in kilograms, and $M$ the reported magnitude number. These scales are equivalent given an ejecta density of 1000 kg m$^{-3}$, and vary by less than half a scale index at other densities. M4/VEI4 corresponds to almost annual events, M5/VEI5 to events such as El Chichón which historically occur every 5 years (Mason et al. 2004), M6/VEI6 to events every 30 years such as Mt. Pinatubo, M7/VEI7 to bicentennial events such as Krakatau, and M8/VEI8 to events rare enough to have not occurred in human history. Non-explosive volcanic events fall below M3/VEI3, since low viscosities lead to mass or volume release in more of a pattern of slow continuous leakage. M3/VEI3 is considered the cutoff for gases to reach the stratosphere.

The eruption of El Chichón in Mexico in 1982 released 7-8 Mt of $SO_2$, and overall represents about one-half of a Pinatubo class event. Six such eruptions have occurred since 1900: Ksudash, Bezymianny, Agung, St. Helens, and Hudson, with one overlap in time. Two other Pinatubo class eruptions have occurred since 1900: Santa Maria in 1902, and Katmai-Novarupta in 1912. While meteorological records exist for these events, no satellite measurements of bulk stratospheric $SO_2$ injections were recorded.

The $SO_2$ input from larger eruptions can only be guessed based on scaling laws. The 1883 eruption of Krakatau is estimated to have released 30-50 Mt of $SO_2$ to the stratosphere, about twice Mt. Pinatubo. Magnitudes and frequencies of larger eruptions are listed in Table 2. The largest known single volcanic event recorded in Earth history is the 27.8 million year old M9.1-9.2 Fish Canyon Tuff deposit in Colorado. Magnitude 9.5-10 eruptions occur effectively only once in Earth history, and will thus be vanishingly rare, at probabilities lower than 1 per billion.

The probability of detecting an eruption of a given magnitude during one year is the same as the annual frequency. For a 30 day nominal residence time in the atmosphere of the $SO_2$ signature, only 12 telescope observations spaced 30 days apart effectively cover the year to detect such a feature in the emergent spectrum. For transmission spectroscopy, the frequency of transit events places an upper limit on such monitoring.

Assuming all objects have Earth-like activity, we define $N$ here as the number of planets times the number of years of observation time. Observing 100 earth-like planets for 10 years yields N=1000 planet-years of equivalent vigilance. The probability of no detections in all observations is given by $P_{fail} = (1-f)^N$ where $f$ is the eruption frequency per year. The probability of detection is then 1- $P_{fail}$.

Table 2 shows the number of planet-years and subsequent discrete telescope observations needed to achieve a detection probability of 1%, 10% and 90% for several eruption magnitudes: e.g. a 1% probability of detecting a 10x event requires 11 planet-years, while a 10% probability for a 10x event requires 106 planet-years, which may be divided over multiple objects. Table 2 shows that detecting even large and rare events is made feasible through repeat observations or a large number of observable Earth-like planets.

The relevant residence time for $SO_2$ is a function of the spectral detection limit, as $SO_2$ is converted to $H_2SO_4$ aerosols. The 17 Mt baseline for Pinatubo was recorded 2 days after the primary eruption, indicating this threshold is held for at least two days. Self et al. (1999) indicate that aerosoliztion might begin at a rate of around -1.0 to -1.5 Mt $SO_2$ per day, but must be non-linear as several Mt $SO_2$ was still detected 28 days later. Note signals that fall below a detectable limit in less than ~15 days are unlikely to achieve full global distribution, as discussed previously in section 2.3. In Table 2 we apply a signal duration Nd = 2, 30, and 170 days to moderate, large and very large eruptions respectively, depending on how far the initial $SO_2$ lies above an assumed 1x baseline detection threshold. This can be thought of as changing not the probability per year, but the number of telescope observations needed to fully cover one year. These observation counts in Table 2 can be adjusted to other instrument sensitivities by changing the parameter Nd.

The change in $SO_2$ with time leads to uncertainty in the magnitude of the event observed. Frequent re-observations following detection may provide the ability to extrapolate backward to constrain the starting magnitude. Change with time will also help differentiate between explosive and non-explosive activity when detected in emergent spectra.

The recent eruption of Mt. Eyjafjallajökull in Iceland during the submission process for this paper suggests that some large non-explosive volcanoes may be able to either deliver $SO_2$ to the stratosphere, or deliver sufficient $SO_2$ to the troposphere over an extended eruption period as to counter the assumption of rapid tropospheric washout. Full data on the altitude and mass of $SO_2$ delivery is not yet published for this event, but $SO_2$ is being tracked by the NASA Aura Ozone Monitoring Instrument (Yang et al. 2009). Early reports suggest a release rate of only ~3 kilotons $SO_2$ per day to between 2-10km (Burton et al. 2010). Probabilities listed in Table 2 consider traditional Plinean activity only, and could be higher if large-scale semi-continuous tropospheric sources do prove significant.

## 3. MODEL DESCRIPTION

Model Earth spectra are calculated with Exo-P (Kaltenegger & Sasselov 2009), a code based on the Smithsonian Astrophysical Observatory code (Traub & Stier 1976). The spectral line database includes the large HITRAN compilation plus improvements from pre-release material and other sources (Rothman et al. 2009; Yung & DeMore 1999). We model the Earth's spectrum using its spectroscopically most significant molecules: $H_2O$, $O_3$, $O_2$, $CH_4$, $CO_2$, CFC-11, CFC-12, $NO_2$, $HNO_3$, $N_2$, and $N_2O$, where $N_2$ is included for its role as a Rayleigh scattering species. Aerosol and Rayleigh scattering are approximated by applying empirical wavelength power laws (Cox 2000) that exert an appreciable effect in the visible blue wavelength range.

Atmospheres from 0 to 100 km altitude are constructed from standard models that are discretized to appropriate atmospheric layers. The area of the stellar disk that is blocked by a transiting planet is given by $\pi R^2(\lambda)$, where $R(\lambda) = R_p + h(\lambda)$, $R_p$ is the radius of the planet at the base of the atmosphere, and $h(\lambda)$ is the effectively opaque height of the atmosphere at that wavelength. $h(\lambda)$ can be as large as about 50 km for strong spectral features in the Earth's atmosphere (see also Kaltenegger & Traub 2009). The fraction of the star's area that is blocked by the absorbing atmosphere, at a given wavelength, is $f_p(\lambda)$ where

$$f_p(\lambda) = 2\pi R_p h(\lambda) / \pi R_s^2 = 2R_p h(\lambda)/R_s^2 \quad (1)$$

The SNR for detecting an atmospheric spectral feature, in an ideal case, is calculated as follows. During a transit, the relevant signal is $N(sig) = N(cont) - N(line)$, where $N(cont)$ is the number of potentially detectable photons in the interpolated continuum at the location of a spectral feature and over the wavelength band of that feature, and $N(line)$ is the number of detected photons in the band. In other words, it is the number of missing photons in the equivalent width of the feature.

The total number of photons detected from the star, in a given spectral range, for transmission spectroscopy is $N_T(sig) = N(tot)*f_p$. N(tot) for emergent spectroscopy is $N_E(sig) = N(tot)*f_S$ where $f_S$ is the suppression factor of the stellar light using e.g. coronagraphy or interferometry in the respective wavelengths.

The noise N(noise) is the fluctuation in the total number of detected photons $N(tot)$ in the same wavelength band, so $N(noise) = N^{1/2}(tot)$, ignoring all other noise contributions. Thus the SNR for detecting a given spectral feature is $SNR = N^{1/2}(tot) * f$, where $f$ is $f_p$ or $f_s$ for transmission and emergent spectroscopy respectively.

The SNR will be dominated by the noise in the noisier of these two spectra (star, or star+planet). The observation time quoted in Table 3 to table assume that one has already determined the stellar spectrum to the required accuracy with either on of two techniques. 1) One can revisit the stellar system to add stellar data during secondary eclipse measurements to characterize the star to the required noise level. This will require the same amount of total time as the planet+star measurement, but will be segmented due to the duration of the secondary eclipse. 2) If the star is quiet/non-active to the level required to measure the planetary atmosphere signatures (see e.g. recent Kepler results, that suggest that 10-20% of stars are quiet to that level (Basri et al. 2010), then one can use one secondary eclipse measurement of the star-only spectrum to achieve the same noise level on both spectra by extrapolating the stellar data, considering only Poisson noise).

Clouds are represented by inserting continuum-absorbing/emitting layers at appropriate altitudes, and broken clouds are represented by a weighted sum of spectra using different cloud layers. We assume that the light paths through the atmosphere can be approximated by incident parallel rays, bent by refraction as they pass through the atmosphere, and either transmitted or lost from the beam by single scattering or absorption. Our line-by-line radiative transfer code for the Earth has been validated by comparison to observed reflection, emission and transmission spectra (Woolf et al. 2002; Turnbull et al. 2006; Christensen & Pearl 1997, Kaltenegger et al. 2007, Irion et al. 2002, Kaltenegger & Traub 2009).

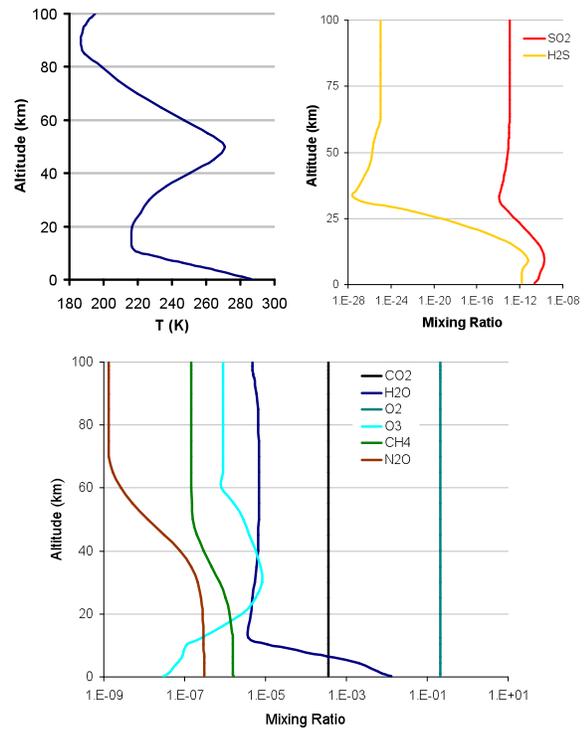

**Figure 1:** (top) Temperature profile from US Standard Atmosphere 1976 (spring/autumn) (COESA 1976) and mixing ratios versus height for the major detectable atmospheric gases up to 100 km for the current Earth (bottom). The current mixing ratios for $SO_2$ and $H_2O$ represent non-eruption levels.

The non-overlapping amounts of an overall 60% opaque water-cloud cover at each altitude are 24% at 1 km, 24% at 6 km, and 12% at 12 km for today's Earth. The effect of shadowing of clouds is taken into account implicitly because the cloud fractions are deduced from modeling the disk-integrated emission spectra of the Earth. A large part of the signal from a disk integrated emission and reflection spectra originates near the limb. Furthermore cloud systems tend to be spaced so that we do not see the cumulative effect of many systems along a given single ray, as we would if a large number of small clouds were scattered uniformly over all of the area.

For this paper we use an Earth model atmosphere (see Fig.1), namely the US Standard Atmosphere 1976 (COESA 1976) spring-fall model, which was verified by observations (Kaltenegger & Traub 09), to derive the strength of detectable features. Because of this input, and in particular the presence of an ozone layer and a temperature inversion layer, planets with differing atmospheric profiles would be expected to have differing results (see e.g. Kaltenegger & Sasselov 2009, Kaltenegger et al. 2007).

## 4. RESULTS

We generate emergent and transmission spectra from 0.4 μm to 40 μm to assess potentially detectable features of volcanism on extrasolar planets using distribution Case 1 and Case 2. The input atmospheric profiles are shown in figure 1. The model spectra are shown in figure 2 and figure 3 (without and with clouds respectively) for Cases 1 and Case 2, for eruptions of 1x, 10x and 100x Mount Pinatubo. Figure 4 and figure 5 show spectra for clouds at varying altitudes for transmission spectra and emergent spectra respectively. Figure 2 and figure 3 show the effective height of atmospheric features in transmission $h(\lambda)$. Figure 3 and figure 5 show the features relative to the signal from a Black Body at 287.9K. Figure 4 shows the transmission feature strength relative to the signal $h_{100km}$ generated by an opaque 100 km high absorption feature in the atmosphere of the transiting body.

The fundamental vibration frequencies shown in figures 2 to 5 are 8.68 μm, 19.30 μm, 7.34 μm, and 3.82 μm for $SO_2$ sorted by their strength. $H_2S$, shows features at 8.45 μm and 3.80 μm, placing them in the mid IR. We only show calculations from 4 to 25 μm instead of 0.4 μm to 40 μm because we concentrate on detectable features of volcanism. Additional features seen in figures 2 to 5, in low resolution, are the 9.6 μm $O_3$ band, the 15 μm $CO_2$ band, the 6.3 μm $H_2O$ band or its rotational band that extends from 12 μm out into the microwave region, a methane feature at 7.66 μm, and three $N_2O$ features at 7.75 μm, 8.52 μm, and 16.89 μm.

Case 2, where volcanic gases are deposited in a layer from 12-25km, shows a stronger feature in the transmission spectra than Case 1 because the material is distributed through more layers, which increases the effect on a transmission spectrum. Figure 2 and figure 3 show that the emergent spectrum is not affected by the altitude distribution of volcanic material because the overall optical depth in the column is independent of the distribution of the gas for an optical thin atmosphere. Part of the purpose of testing Case 1 was to see if highly concentrating $SO_2$ could improve detectability, however results show that it is the broader distribution of a Pinatubo-like Case 2 which increases the detectable signal in transmission. Table 3 shows the SNR for primary eclipse transit observations, per transit, and for a set time of 100hrs, what translates to about 7 transits. Table 4 shows the SNR for secondary eclipse observations, per transit, and per 170 days, the $SO_2$ residence time in the stratosphere. Both tables list data for the closest G star α Centauri A, and a G star at a distance of 10pc. These idealized results of photon noise limited calculations can be used as input to different models of instrument noise in individual JWST instruments and observation strategies. Detectable atmospheric features, like the SO$_2$ features described here to assess volcanism, that are not present in high quantities in Earth's atmosphere are excellent features to look for in exoplanet atmospheres from future Ground-based telescopes like ELT.

converted to assess what observation time is needed to observe such features from the ground – depending on the telescope size, instrument characteristics as well as observation site quality as function of wavelength.

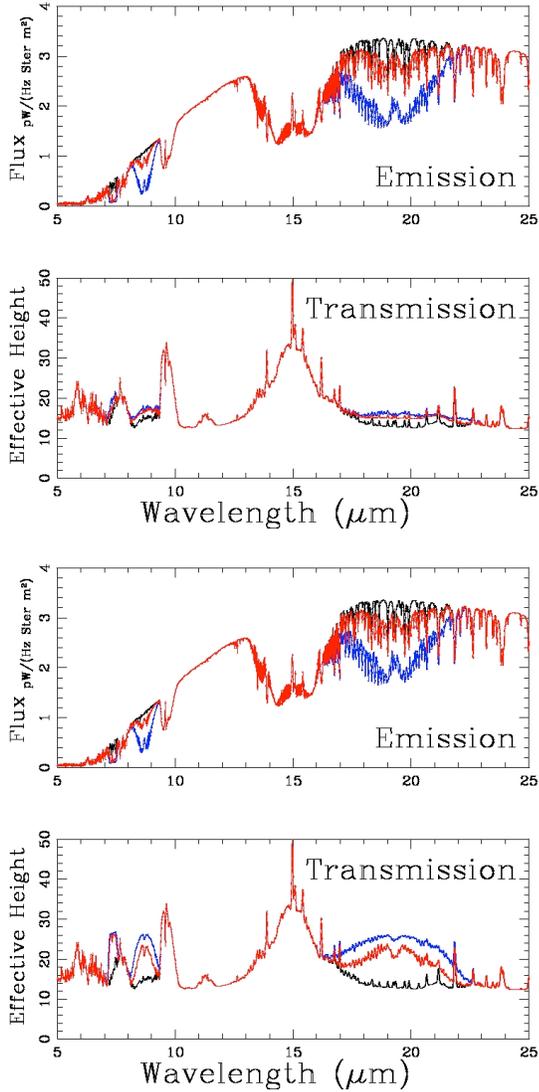

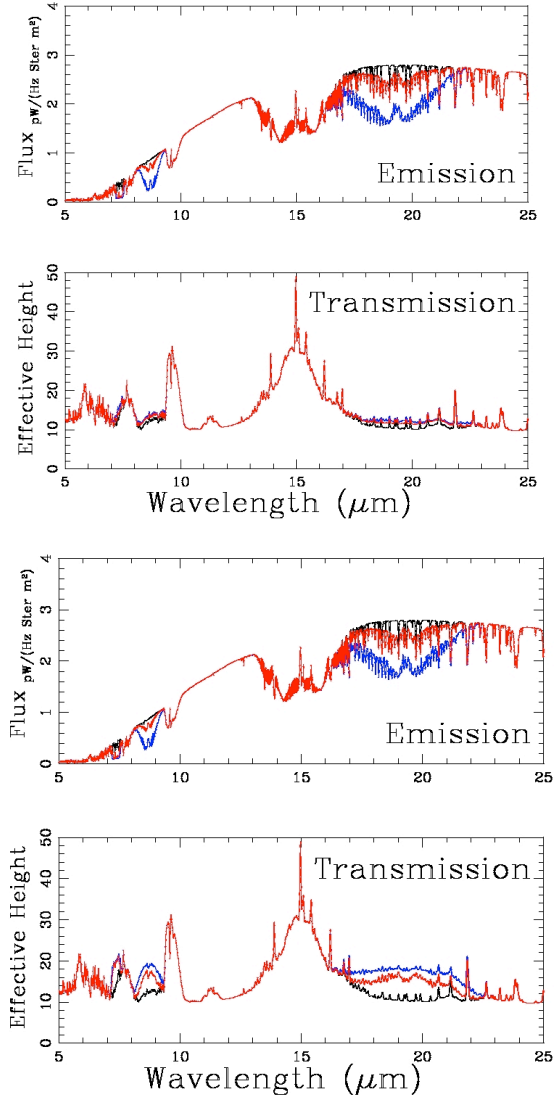

**Figure 2:** Case 1, thin layer (top). Case 2, thick layer (bottom). No clouds. Relative flux. Model emergent spectra 0.4 µm – 40 µm of a cloud-free atmosphere for three volcanic SO$_2$ concentrations (Black: No eruption. Red: 10x Pinatubo eruption (baseline). Blue: 100x baseline). The calculations are performed at very high resolution (0.1 wavenumbers) and subsequently smoothed for display (R = 150). Three sulfur dioxide features become detectable in the 10x and 100x cases.

The values in table 3 to table 6 can easily

**Figure 3:** Results with clouds. Absolute values. Case 1, thin layer (top). Case 2, thick layer (bottom). Model emergent spectra 5 µm – 25 µm of an Earth atmosphere including water clouds at 1km, 5km and 12 km for three volcanic SO$_2$ concentrations (Black: No eruption. Red: 10x Pinatubo eruption (baseline). Blue: 100x baseline). The calculations are performed at very high resolution (0.1 wavenumbers) and subsequently smoothed for display (R = 150). SO$_2$ feature visibility is similar to the no cloud case.

Figure 4 and figure 5 show the effect of clouds at varied altitudes.

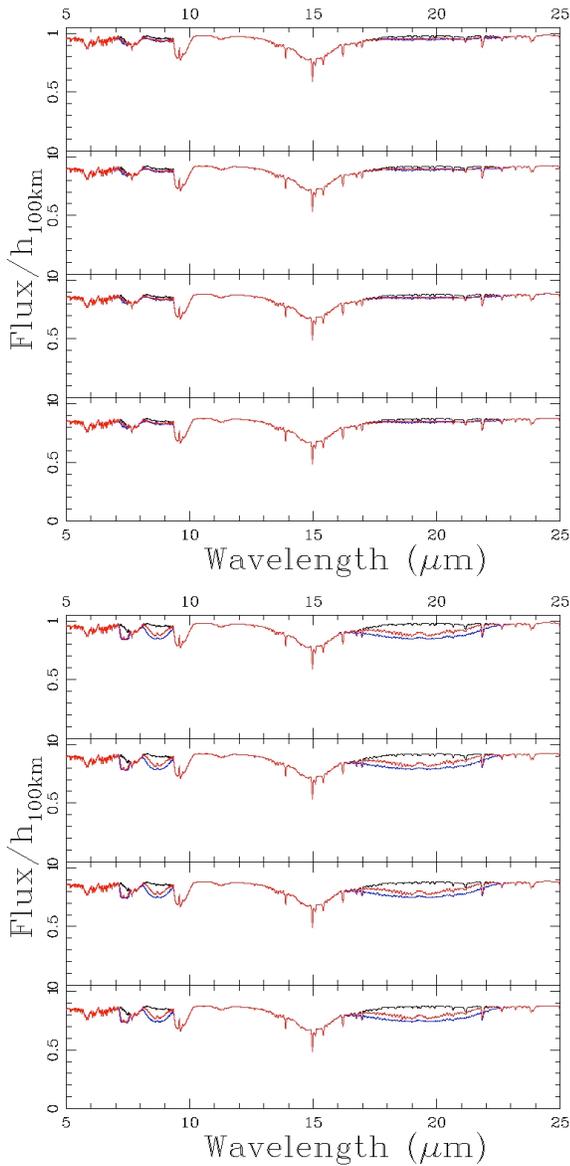

Figure 4: Panels top to bottom: 60% overall cloud cover at 12km, 5km, 1km, and no-cloud model , 5 µm – 40 µm of an Earth atmosphere for three volcanic $SO_2$ concentrations (Black: No eruption. Red: 10x Pinatubo eruption (baseline). Blue: 100x baseline). Relative flux shown. Top: Case 1 transmission spectra, Bottom: Case 2 transmission spectra. The calculations are performed at very high resolution (0.1 wavenumbers) and subsequently smoothed for display (R = 150). Note that in transmission, note how cloud altitude has little effect on signal because gases are placed in the stratosphere.

There are no changes in the individual emergent cloud spectra between Cases 1 and 2 because volcanic material is deposited above the uppermost cloud layer on Earth at 12 km. Note how cloud altitude is shown to have a strong impact on overall emergent spectra, but little impact to transmission spectra, because transmission spectra primarily sample above our highest 12km clouddeck, due to the longer optical pathlengths at lower altitudes.

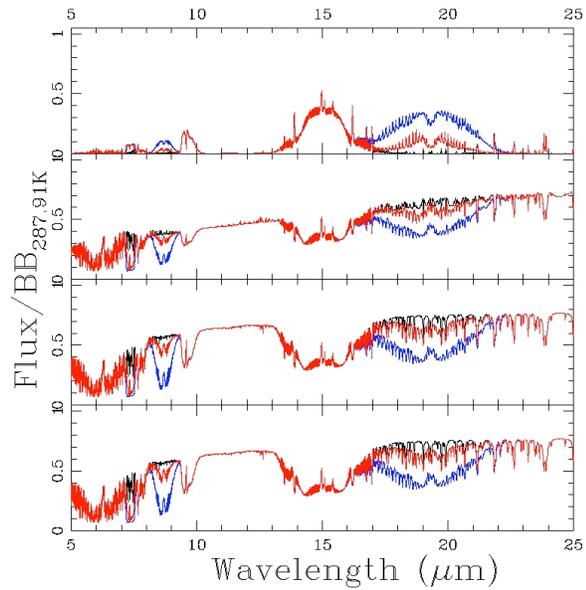

Figure 5: Panels top to bottom: 60% overall cloud cover at 12km, 5km, 1km, and no-cloud model , 5 µm – 40 µm of an Earth atmosphere for three volcanic $SO_2$ concentrations (Black: No eruption. Red: 10x Pinatubo eruption (baseline). Blue: 100x baseline). Relative flux shown. Emergent spectra for both Cases 1 and 2. The calculations are performed at very high resolution (0.1 wavenumbers) and subsequently smoothed for display (R = 150). Note how detectable $SO_2$ features blend with cloud spectra in emission.

In general, volcanic signatures become visible in a present day Earth atmosphere between 1x and 10x the Pinatubo baseline. In transmission, the $SO_2$ feature is not detectable at 1x the baseline, clearly visible at 10x the baseline, and little changed at 100x. This signal saturation effect begins to occur in comparison to the emergent spectrum because of the overall increased optical path of the light through the atmosphere in transmission. Achieving 2x to 10x baseline concentrations can be accomplished by observing a Krakatau to Tambora class event, or due to a combination of smaller eruptions.

## 5. DISCUSSION

### 5.1 Alternative Sulfur Sources

The lifetime of sulfur species that today are rapidly oxidized in Earth's atmosphere is longer under reducing conditions like on the early Earth. Increased volcanic activity and outgassing of $CO_2$, $SO_2$, and $H_2S$ could exhaust the oxidant supply in a reduced atmosphere, allowing $SO_2$ to reach concentrations of several hundred ppm near the planet's surface (see also Kasting et al. 2010, Halevy et al. 2007). The potential buildup of sulfur components as a result of maintained volcanism and the remote detectability of a global sulfur cycle is discussed elsewhere (Kaltenegger & Sasselov 2009). When volcanism subsides, $SO_2$ would be rapidly removed from the atmosphere by continued photolysis and by reaction with oxidants, now supplied at a faster rate than reduced gases. Reducing precipitation would also lead to a build-up of sulfur bearing species in a planet's atmosphere, making an exoplanet with decreased precipitation and increased volcanism an ideal candidate to study for volcanic eruptions.

The only sulfur containing volcanic gases originating in Earth's troposphere which reach the stratosphere are the comparatively inert compounds OCS and $CS_2$, at the relatively small rate of $9.4 \times 10^4$ to $3.1 \times 10^8$ kg/yr (Crutzen 1976), and $1.3 \times 10^4$ to $4.4 \times 10^7$ kg/yr (Halmer et al 2002) respectively. Volcanic sulfur emissions to Earth's troposphere are equivalent to approximately 14% of anthropogenic emissions (Robock 2000). It is conceivable yet unlikely that a continuous $SO_2$ signal on an exoplanet may arise from natural combustion processes such as forest fires or long duration coal seam fires producing sufficient OCS and $CS_2$ to diffuse upwards.

Halmer et al. (2002) estimate the annual $SO_2$ injection to the stratosphere from all sources, including OCS migration and small eruptions is 2.1-3.2 Mt, or ~10x less than a single Pinatubo event. On a dry planet where $SO_2$ is not aerosolized, this background signal may be 2.4-4.0 Mt (if the washout effect in the plume is eliminated) or 15.0-21.0 Mt (if all annual $SO_2$ were able to diffuse upwards). A high annual background would make it difficult to detect an individual eruption.

### 5.2 Increased Volcanism on Young Earths and super-Earths?

The scenario of a highly volcanically active Earth-like planet is based on several effects that especially apply to young, tidally heated, and more massive rocky planets. High volcanism rates may allow non-explosive eruptions to build up daily average $SO_2$ concentrations comparable to explosive events, provided washout effects are limited, and allow non-explosive activity to become remotely detectable (see Kaltenegger & Sasselov 2009).

For rocky super-Earths (rocky planets with up to 10 $M_E$) (Valencia et al. 2007), volcanic activity may be higher due to a smaller surface-to-volume ratio, where radionuclide and accretion heating scale by the total silicate mass fraction and total planet mass respectively. If the probability of a given eruption is considered on a per-surface area basis, then the probability of events given in Table 2 may reasonably be increased by a factor of $R/R_E$, the relative surface-to-volume increase, where R is the super-Earth radius and $R_E$ the radius of Earth. This effect however may be offset by higher global water mass fractions, submerged continents, and thicker atmospheres, all of which can make it harder for volcanic gases to reach observable altitudes. There are multiple competing effects in a super-Earth atmosphere that influence the strength of the absorption features, e.g. the scale height decreases due to increased gravity while the total atmospheric mass and the pathlength of a ray through the limb increases, because the overall volume of the atmosphere changes (see Kaltenegger et al. in prep). The overall rate of volcanism on such worlds, including non-explosive events, is discussed in Kite et al. (2009).

Tidal heating may contribute significantly to volcanism for eccentric exomoons or eccentric planets in heliocentric periods of 20-30 days or less (see Henning et al., 2009 for a detailed discussion). Such tidal heating has the potential to generate from thousands to millions of times more internal heating than in the modern Earth. However, at the upper end of this heat range, magma is more likely to escape in a non-explosive pattern, or may simply emerge into lava lakes or magma oceans partly sustained by the high insolation values near stars, such as the magma ocean suggested for Corot 7b. Such extreme worlds may also be more likely to have a reducing atmosphere. Significant tidal activity can be stimulated in multi-body systems by secular perturbations, secular resonance crossings, or a deep and stable mean motion resonance, analogous to the Galilean moon system. Modest tidal heating cases may easily supply some exoplanets with both a 10x increase in the size and frequency of large eruptions, while simultaneously enhancing non-explosive activity.

Planet age may have a strong effect on volcanic intensity (see e.g. Isley and Abbott, 2002). A young planet is expected to have more residual accretion heat and higher radionuclides, at 5x to 10x more than Earth's modern ~40 terawatt internal heat rate (Turcotte and Schubert, 2002). Very young planets may have increased activity from $^{26}$Al decay, or may still be cooling after large embryonic mergers. Continental configurations also vary with plate tectonic activity, and certain epochs in Earth's history included major subduction events, leading to a significant increase in the frequency of explosive volcanoes. On the other hand, large continents and their corresponding subduction zones were less common on the very early Earth (Condie, 1998), suggesting that more of this early heat would escape in the form of non-explosive mid-ocean ridges and basaltic ocean islands.

It is not clear whether hotter planets would show more large explosive eruptions. More vigorous mantle convection would be expected to result in enhanced mid-ocean ridge, hotspot, and ocean island volcanism, all of which inject gases at or near sea level. Hotter magmas are less viscous, allow greater migration and escape of volatiles and bubbles, and tend to erupt non-explosively. Moderately enhanced convection should increase the occurrence of Large Igneous Province (LIP) events, such as the Siberian and Deccan traps. These may occur when convective plume heads reach the surface. LIP events are associated more with broad fissure fields and flowing magmas than with Plinean volcanism or distant ash deposits, and thus may not contribute to observable stratospheric gas concentrations. However, geologic records, such as those of the Yellowstone, Colorado, and Snake River basins, clearly show M8 level explosive volcanic activity divided into two primary pulses: 36-25 Ma ago and 13.5 Ma ago to present (Mason et al., 2004) a fact likely linked to the rise of mantle plumes or similar structures. Modest heat flow enhancements should also increase subduction rates on Earthlike worlds.

The initial concentration of water on exoplanets is expected to vary widely based on formation patterns and total mass. Terrestrial bodies that either start out substantially drier or become substantially drier may have reduced explosive volcanism, due to lithospheres too viscous for subduction or thick static lids. On planets with high water content, magmas may also be significantly less viscous, reducing explosivity, in addition to global oceans potentially submerging otherwise subaerial volcanoes. Global or local variations in total sulfur content, or changes in total sulfur solution into the iron core, can alter the amount of $SO_2$ released in any given volcanic eruption from those modeled here. Remote observation of volcanic eruptions on exoplanets would help constrain such issues while enriching our current knowledge of the diversity and prevalence of planetary volcanism substantially.

### 5.3 Primary Eclipse – Observation Requirements

Throughout this paper we treat transmission and emergent spectroscopy similarly, but the probability of detecting an eruption of a given magnitude during a time period is significantly lower for a transiting planet if the time between transits is longer than the residence time of volcanic gases in the planet's atmosphere (e.g. 170 days). This makes such observations primarily feasible for planets closer to their host stars.

Unlike emergent spectra that can be observed continuously by direct imaging and secondary eclipse measurements, the telescope integration time for transmission is additionally limited by transit duration. Durations are about 12.9 hrs for an Earth at 1 AU and less for closer objects.

To detect the strongest $SO_2$ feature at 19.5 μm for Case 2 with a SNR of 3 per individual observation, assuming an efficiency of 0.15 and only considering photon noise, would require a collecting area of 225 m$^2$ for an Earth-like planet around a star at 10pc, and 63 m$^2$ for the closest G star α Centauri A to be feasible. With an efficiency of 0.3 these values change to 115 m$^2$ and 32 m$^2$ respectively. Note that JWST has a collecting area of 33.18 m$^2$.

Generally the SNR for a transiting planet for a set amount of time increases with a decreasing planet-to-star contrast ratio, making M stars a good target star choice. The shorter orbital period for planets around M stars can is helpful, if an eruptive signal endures through several transits, allowing for increased overall integration times, on the other hand, the transit duration for a planet orbiting an M star is shorter (see Kaltenegger & Traub in prep for details). The SNR scales with the radius of the planet and inverse with the area of the star (see equ.1). Table 5 illustrates how the SNR could increase for smaller host stars (we choose a M2 V star with 0.44 solar radius and a temperature of 3400 K) and bigger planets (1.5 Earth radii). Note that this example only illustrates the effects.

The atmospheric profile of the Super-Earth has not been calculated self consistently but the absorption features are assumed to be Earth-analog in first approximation (see Kaltenegger & Traub in prep. for details). Primary Eclipse SNR for a M2V host star and a Super-Earth with 1.5 Earth radii are given in table 5. The SNR per transit is comparable at a set stellar distance (e.g. 10pc), even so the observation time per transit is shorter by about a factor of 10 (transit duration of a planet in the HZ is about 12 hrs for the Sun and about 1.7 hrs for a M2 star). For a set observation time of 100hrs, what translates in about 7 transits for a G star and about 60 transits for a G2V star, the SNR for the small host star increases dramatically. We use Gl 144 as the closest M2V star at 2.45 pc. Transit spectroscopy observations would be an interesting way to probe the distribution of gases as a function of height in the atmosphere, but this level of accuracy does not appear feasible in the near future. Generally, a combination of observations of emergent and transmission spectra would allow for an in-depth characterization of a volcanic planet.

## 5.4 Secondary Eclipse Measurements and Direct Imaging – Observation Time

To detect the strongest $SO_2$ feature in secondary eclipse or direct imaging, we calculate the SNR and observation times shown in table 4. To observe the strongest $SO_2$ feature at 19.5 μm for Case 2 with a SNR of 3 per individual observation, assuming an efficiency of 0.15, photon noise, and JWST's collecting area of 33.18 m$^2$, would require an observation time of 316.5 days for an Earth-like planet around a star at 10 pc, and 5.8 days for α Centauri A in secondary eclipse measurements, assuming only photon noise from the star, which is assumed to be characterized to such a level of accuracy. With an efficiency of 0.3 these values change to 158.1 days and 2.8 days respectively. Secondary eclipse observations are not limited by short transit times as primary eclipse observations are, since photons from the planet are collected over more of the orbital period. Table 6 illustrates how the SNR could increase for smaller host stars (we choose an M2 V star with a temperature of 3400 K) and bigger planets (2 Earth radii). Note that this example only illustrates the effects. The atmospheric profile of the Super-Earth has not been calculated self consistently but is assumed to be Earth-analog in first approximation. Secondary Eclipse SNR scale with the area of the planet and inverse with the area of the star. SNR for a M2 V host star with a 0.44 solar radius, 3400 K and a Super-Earth with 1.5 Earth radii are given in table 6. The SNR increases by a factor of about 2 per transit at a set stellar distance (e.g. 10pc) even so the observation time per transit is shorter by about a factor of 10 (see column per transit). A set observation time of 300 hrs for the closest M2V star yields about the same SNR as a 170 day observation for the closest Sun-like star. Especially for secondary eclipse measurements the SNR for the small host star increases drastically, making small stars the best targets. We use Gl 144 as the closest M2V star at 2.45 pc.

To compare these results to planned direct imaging missions that suppress the starlight while collecting light from the planet itself, one increases the SNR by the square root of the background suppression factor (of mainly the starlight and zodiacal dust, both in our system and the exoplanet system).

Section 1.1 discussed that emergent spectra can show $SO_2$ from all types of volcanism, including sea level sources. Thus probabilities for a detection of either origin, explosive or non-explosive, are better than values listed in Table 2. This combined with the observation requirements above, will make emergent spectra a highly useful tool for volcanic exoplanet studies.

## 5.5 Repeat Observations - Localizing a Source – Far Future

Multiple high SNR observations of an exoplanet with a putative volcanic signature over several months to years following an initial signal detection, could greatly increase our understanding of the object and the type of volcanism occurring. This would be the first test of the $SO_2$ residence time baseline versus the atmospheric composition – reducing versus oxidizing. Re-observations would help distinguish if a detection is from a semi-continuous source, such as Eyjafjallajökull, which spanned two full years in its most recent previous eruption, 1821-1823 (Sturkell et al. 2010). If an $SO_2$ signal remains past this timescale, it suggests the origin is either multiple smaller eruptions whose signal is combining over time, a stratospheric chemistry different from Earth's, or a high rate of large eruptions. A signal disappearing in under 170 days would suggest that the initial discovery was of a single quantifiable volcanic event.. Detailed observation records and the concurrent detection of other chemicals, including water, might eventually allow for the classification of a volcanic source region and tectonic setting. In rare cases, a transit may be able to detect an eruptive plume while it is still localized over the source volcano that is located on the limb of the planet. In one day, (June 15, 5pm to June 16, 5pm) the main plume from Mt. Pinatubo spread roughly 15 degrees in both latitude and longitude (Self et al. 1996). If the target planet's rotation rate, obliquity axis, transit duration, and transit frequency were each favorable, and an observed eruption were still partly localized, then $SO_2$ signature variation by limb longitude could potentially be used to help constrain the planet's rotation rate and to begin mapping the surface relative to future eruptive events. This technique would require a high frequency of repeat observations as well as a very high SNR per observation, which is potentially possible with second generation future space missions.

## 6. CONCLUSIONS

We show that large-scale explosive volcanism can be remotely observed on exoplanets around the closest stars using sulfur dioxide as a chemical proxy. We have developed two models for present day Earth-like planets, with Case 1 depositing volcanic gases into a 2km thick stratospheric layer, and Case 2 depositing gases between 12 and 25km in altitude. We have derived the observable features for primary and secondary eclipse measurements and discussed the likelihood of observing volcanism on Earth to super-Earth exoplanets.

Table 3 and 4 show the SNR per transit and the observation time needed, respectively, for secondary eclipse measurements to detect explosive volcanism on exoplanets using a 6.5m telescope in space such as JWST. Secondary eclipse measurements with JWST could detect the signature of volcanism on exoplanets due to the increased available observation time. Primary eclipse measurements are limited to the closest stars due to the limiting transit duration time. In our calculations we have only included photon noise from the star. The idealized results from these photon noise limited calculations can be used as input to the different models for instrument noise of individual JWST instruments and observation strategies. Detectable atmospheric features, like the $SO_2$ features described here to assess volcanism, that are not present in high quantities in Earth's atmosphere are excellent features to look for in exoplanet atmospheres from future Ground-based telescopes like ELT and can easily be determined from Table 3 to 6 depending on the instrument, observation site and telescope size.

Volcanic features generally become visible between 1-10 times the Mt. Pinatubo eruption. Case 2 shows a stronger feature in the transmission spectra than Case 1 because volcanic material is distributed through more layers. We show that the probability to observe a Pinatubo class explosive eruption is about 1% if 4 Earth-like planets are observed for 1 year. A higher contrast signal from a 10x event can be detected with 10% probability by any equivalent to observing ~50 Earth-like planets for 2 years. Planet geology suggests that probabilities are higher for young hot planets, tidally heated bodies, and continent bearing super-Earths. Tropospheric $SO_2$ distributions are rapidly aerosolized in the presence of water and were not modeled, but may reach levels detectable in emergent spectra on such worlds with increased volcanism. These observations becomes a very interesting option to characterize rocky planets, especially if one assumes larger, and/or more frequent eruptions than on Earth, or smaller host stars, where a planet in the HZ orbits closer to their stars, increasing the transit probability., or longer $SO_2$ residence times than on Earth. Table 5 and table 6 illustrate how the SNR could increase for smaller host stars and Super-Earths, assuming an Earth-analogue atmosphere in first approximation. Note that the atmospheric profile of the Super-Earth has not been calculated self-consistently here. We use this case to illustrate the strength of this approach for small host stars with Super-Earths. For both primary and especially for secondary eclipse measurements the SNR for the small host star and bigger planets increase drastically, making small stars the best targets for a search with JWST. All these hypotheses can be tested by remotely observing exoplanets in low resolution and scanning them for features that can be attributed to volcanism. Remote observation of volcanic eruptions on exoplanets would enrich our current knowledge of the diversity and prevalence of volcanism substantially.

**Acknowledgements**

LK, WH & DS gratefully acknowledge the Origins of Life Initiative at Harvard University and the NASA Astrobiology Institute. This work has also been supported by NSF Grant EAR0440017. We thank Jeff Standish and Richard J. O'Connell for helpful reviews and Jean-Michel Desert and David Charbonneau for helpful discussions.

**Table 1. Composition of Volcanic Gases**

| Volcanic Gas Species | Volume, Ave. Subduction Zone (mol%) | Volume, Ave. Rift Zone (mol%) | Mt. Pinatubo (1991), Total Mass (Mt) |
|---|---|---|---|
| $H_2O$ | 95.4 | 63.6 | 250-500 † |
| $CO_2$ | 6.37 | 20.6 | 82-104 ‡ |
| $SO_2$ | 3.72 | 9.49 | 15-19 |
| $H_2S$ | 1.55 | 0.91 | 1.2-1.6 ‡ |
| $H_2$ | 1.19 | 4.91 | 0.2-0.5 ‡ |
| HCl | 0.983 | - | 0-3 |
| HF | 0.097 | - | - |
| CO | 0.024 | 0.92 | - |
| OCS | 0.014 | 0.00071 | - |

*Notes: Adapted from Halmer et al. 2002 and Self et al. 1996. Columns 1 and 2: Average composition of volcanic gases from 13 subduction zone volcanoes compared to 4 rift zone volcanoes, in mol%. See Halmer et al. 2002 for individual references. Certain volcanoes known to be Cl or S rich outliners such as Mt. Augustine or Mt. Etna are excluded. CO, HF, OCS values based on smaller numbers of measurements. Column 3: Absolute amounts from the 1991 Mt. Pinatubo eruption, in metric megatons (Mt). † Estimate based on pumice gas inclusion measurements. ‡ Estimates based on average subduction zone volcanic gas ratios and measured Pinatubo $SO_2$ amounts.*

**Table 2. Frequency of Explosive Volcanic Events**

| Name | | El Chichón | Pinatubo | Krakatau | Tambora | Taupo | Toba / Yellowstone |
|---|---|---|---|---|---|---|---|
| ~ x Baseline | | 0.5x | 1x | 2x | 10x | 100x | 500 / 1000x |
| Year | | 1982 | 1991 | 1883 | 1815 | 23500 y.a. | 73000 / $6e^5$ y.a. |
| VEI / Mag. | | 5 | 6 | 7 | 7 | 8.1* | 8.8* / 8.7-8.9* |
| Stratospheric $SO_2$ (Mt) | | 7-8 | 17 | 30-50 | ~200 est. | ~2000 est. | ~$1e^4$ / $2e^4$ est. |
| Frequency Estimate, f (1/yr) | | 0.05-0.2** | 0.03 | 0.002 | 0.001 | $1e^{-4}$-$1e^{-6}$** | $1e^{-6}$-$1e^{-8}$** |
| Planet-years to achieve P = | 1% | 1 | 1 | 6 | 11 | $1e^2$-$1e^4$ | $1e^5$-$1e^6$ |
| | 10% | 1-3 | 4 | 53 | 106 | $1e^3$-$1e^5$ | $1e^6$-$1e^7$ |
| | 90% | 11-45 | 76 | 1151 | 2302 | $2e^4$-$2e^6$ | $2e^7$-$2e^8$ |
| Signal Duration, Nd (days) | | n/a | 2 | 30 | 170 | 170 | 170 |
| Observations to achieve P = | 1% | n/a | 183 | 73 | 24 | $2e^2$-$2e^4$ | $2e^5$-$2e^6$ |
| | 10% | n/a | 730 | 645 | 228 | $2e^3$-$2e^5$ | $2e^6$-$2e^7$ |
| | 90% | n/a | 13870 | 14003 | 4943 | $5e^5$-$5e^6$ | $5e^7$-$5e^8$ |

*Notes: Frequency comparison of various magnitude volcanic events to the 1991 Pinatubo baseline adopted in this paper. Probabilities assume Earthlike behavior, and may be higher for younger or more tidally active planets. $SO_2$ to the stratosphere for events larger than Krakatau are highly uncertain. For Toba and Yellowstone, magnitudes are well known but frequencies are highly uncertain and span the same effective range for both events. Note the smaller number of observations needed for a Tambora-class 10x event due to a detectable signal surviving longer in the atmosphere. Observations to achieve P assume minimum detectability is 1x baseline.* Magnitudes directly measured from the mass of ejecta deposits. ** Frequencies extrapolated from the scaling laws and historical evidence presented in Mason et al. 2004.*

**Table 3. Primary Eclipse Signal to Noise Ratios**

| 6.5-m telescope | | | | SNR transit$^{-1}$ | | SNR 100hr$^{-1}$ (7.2 transits) | |
|---|---|---|---|---|---|---|---|
| Feature | λ(μm) | Δλ(μm) | h(λ),km | 10pc | 1.34pc | 10pc | 1.34pc |
| $H_2O$ | 6.3 | 1.0 | 7 | 0.33 | 2.5 | 0.92 | 6.9 |
| $CH_4$ | 7.7 | 0.7 | 7 | 0.20 | 1.5 | 0.56 | 4.2 |
| $O_3$ | 9.8 | 0.7 | 30 | 0.61 | 4.5 | 1.69 | 12.6 |
| $CO_2$ | 15.2 | 3.0 | 25 | 0.58 | 4.3 | 1.60 | 12.0 |
| **Case 2** | | | | | | | |
| $SO_2$ | 7.5 | 0.5 | 5 | 0.13 | 0.9 | 0.35 | 2.6 |
| $SO_2$ | 9 | 1 | 7 | 0.14 | 1.0 | 0.38 | 2.9 |
| $SO_2$ | 19 | 4 | 10 | 0.19 | 1.5 | 0.54 | 4.0 |

*Notes: Major spectroscopic features in transmission (col. 1-4) and SNR of a transiting Earth, per transit and per 100hr, and the number of transits needed for a 6.5-m space based telescope, for the closest G star and the Sun at 10 pc, assuming an efficiency of 0.15 for a 100x Pinatubo eruption.*

**Table 4. Secondary Eclipse Signal to Noise Ratios**

| 6.5-m telescope | | | | SNR transit$^{-1}$ | | SNR 170 days | |
|---|---|---|---|---|---|---|---|
| Feature | λ(μm) | Δλ(μm) | Equ.w.(λ) | 10pc | 1.34pc | 10pc | 1.34pc |
| $H_2O$ | 6.3 | 1.0 | 0.27 | 0.004 | 0.03 | 0.08 | 0.59 |
| $CH_4$ | 7.7 | 0.2 | 0.02 | 0.003 | 0.02 | 0.05 | 0.37 |
| $O_3$ | 9.8 | 0.5 | 0.08 | 0.014 | 0.11 | 0.25 | 1.90 |
| $CO_2$ | 15.2 | 3.0 | 0.52 | 0.093 | 0.70 | 1.65 | 12.36 |
| **Case 1 = Case 2** | | | | | | | |
| $SO_2$ | 7.5 | 0.1 | 0.05 | 0.006 | 0.05 | 0.11 | 0.84 |
| $SO_2$ | 9 | 2 | 0.23 | 0.021 | 0.16 | 0.37 | 2.78 |
| $SO_2$ | 19 | 4 | 0.34 | 0.124 | 0.93 | 2.20 | 16.44 |

*Notes: Major spectroscopic features in secondary eclipse (emergent spectrum) (col. 1-4) and SNR of an Earth, per transit and per 170 days (4080 hr) for a 6.5-m space based telescope, for the closest G star and the Sun at 10 pc, assuming an efficiency of 0.15 for a 100x Pinatubo eruption.*

**Table 5. Primary Eclipse Signal to Noise Ratios: M2 V host star and a Super-Earth with 1.5 Earth radii**
*\* This is an illustration how the SNR increases for smaller host stars and bigger planets. The atmospheric profile of the Super-Earth has not been calculated self consistently but is assumed to be Earth-analog in first approximation.*

| | 6.5-m telescope | | | SNR transit$^{-1}$ | | SNR 100hr$^{-1}$ (60 transits) | |
|---|---|---|---|---|---|---|---|
| Feature | $\lambda(\mu m)$ | $\Delta\lambda(\mu m)$ | $h(\lambda)$,km | 10pc | 2.45pc | 10pc | 2.45pc |
| $H_2O$ | 6.3 | 1.0 | 7 | 0.29 | 1.1 | 2.2 | 8.7 |
| $CH_4$ | 7.7 | 0.7 | 7 | 0.18 | 0.7 | 1.4 | 5.4 |
| $O_3$ | 9.8 | 0.7 | 30 | 0.55 | 2.2 | 4.2 | 16.6 |
| $CO_2$ | 15.2 | 3.0 | 25 | 0.53 | 2.1 | 4.1 | 16.0 |
| **Case 2** | | | | | | | |
| $SO_2$ | 7.5 | 0.5 | 5 | 0.11 | 0.4 | 0.9 | 3.4 |
| $SO_2$ | 9 | 1 | 7 | 0.12 | 0.5 | 0.9 | 3.7 |
| $SO_2$ | 19 | 4 | 10 | 0.18 | 0.7 | 1.4 | 5.4 |

*Notes: Major spectroscopic features in transmission (col. 1-4) and SNR of a transiting Super-Earth, per transit and per 100hr, and the number of transits needed for a 6.5-m space based telescope, for the closest M2V star and the M2V star at 10 pc, assuming an efficiency of 0.15 for a 100x Pinatubo eruption. We use Gl 144 as the example for the closest M2V star.*

**Table 6. Secondary Eclipse Signal to Noise Ratios: M2 V host star and a Super-Earth with 1.5 Earth radii**
*\* This is an illustration how the SNR increases for smaller host stars and bigger planets. The atmospheric profile of the Super-Earth has not been calculated self consistently but is assumed to be Earth-analog in first approximation*

| | 6.5-m telescope | | | SNR transit$^{-1}$ | | SNR 100 hr$^{-1}$ | |
|---|---|---|---|---|---|---|---|
| Feature | ●$(\mu m)$ | $\Delta\lambda(\mu m)$ | Equ.w.$(\lambda)$ | 10pc | 2.45pc | 10pc | 2.45pc |
| $H_2O$ | 6.3 | 1.0 | 0.27 | 0.011 | 0.05 | 0.09 | 0.35 |
| $CH_4$ | 7.7 | 0.2 | 0.02 | 0.007 | 0.03 | 0.06 | 0.22 |
| $O_3$ | 9.8 | 0.5 | 0.08 | 0.036 | 0.14 | 0.28 | 1.10 |
| $CO_2$ | 15.2 | 3.0 | 0.52 | 0.231 | 0.91 | 1.78 | 7.00 |
| **Case 1 = Case 2** | | | | | | | |
| $SO_2$ | 7.5 | 0.1 | 0.05 | 0.016 | 0.06 | 0.12 | 0.49 |
| $SO_2$ | 9 | 2 | 0.23 | 0.053 | 0.21 | 0.41 | 1.61 |
| $SO_2$ | 19 | 4 | 0.34 | 0.305 | 1.20 | 2.35 | 9.25 |

*Notes: Major spectroscopic features in secondary eclipse (emergent spectrum) (col. 1-4) and SNR of an Earth, per transit and per 100hrs for a 6.5-m space based telescope, for the closest G star and the Sun at 10 pc, assuming an efficiency of 0.15 for a 100x Pinatubo eruption. We use Gl 144 as the example for the closest M2V star.*